\documentclass[%
 aip,
 amsmath,amssymb,
 reprint,%
]{revtex4-1}

\usepackage{graphicx}% Include figure files
\usepackage{dcolumn}% Align table columns on decimal point
\usepackage{bm}% bold math
\usepackage[utf8]{inputenc}
\usepackage[T1]{fontenc}
\usepackage{mathptmx}
\usepackage{etoolbox}
\usepackage{hyperref}
\usepackage{dsfont}
\usepackage{physics}
\usepackage{siunitx}
\makeatletter
\def\@email#1#2{%
 \endgroup
 \patchcmd{\titleblock@produce}
  {\frontmatter@RRAPformat}
  {\frontmatter@RRAPformat{\produce@RRAP{*#1\href{mailto:#2}{#2}}}\frontmatter@RRAPformat}
  {}{}
}%
\makeatother
\begin{document}

\preprint{AIP/123-QED}

\title{Optical sensing of charge and spin current fluctuations in centrosymmetric semiconductors}
% Force line breaks with \\
\author{Amin Lakhal}
\affiliation{femtoQ Laboratory, Department of Engineering Physics, Polytechnique Montréal, Montréal QC, Canada}
\author{Stéphane Virally}
\affiliation{femtoQ Laboratory, Department of Engineering Physics, Polytechnique Montréal, Montréal QC, Canada}
\author{Jacob B. Khurgin}%
\affiliation{Electrical and Computer Engineering Department, Johns Hopkins University, Baltimore, MD 21218, United States}
\author{Denis V. Seletskiy}
\email{denis.seletskiy@polymtl.ca}
\affiliation{femtoQ Laboratory, Department of Engineering Physics, Polytechnique Montréal, Montréal QC, Canada}
\date{\today}% It is always \today, today,
             %  but any date may be explicitly specified

\begin{abstract}
We propose a time-resolved optical measurement scheme for sampling transient charge and spin currents in a bulk centrosymmetric semiconductor. The technique relies on emission of second harmonic light triggered by  
%(Jacob)I would not call it FWM. It is more of EFISH but biased by curent not field 
a pulsed below-gap optical excitation and a spontaneous intraband polarization arising from spin or charge motion, mediated by a $\chi^(3)$-based nonlinear optical process. Our proposal uses homodyne amplification to boost the weak second harmonic signal, making it detectable with conventional electronics, calculated for charge current in a room temperature GaAs semiconductor.  
This all-optical technique requires neither electrical contact nor bias fields and the signal is estimated at a few percent relative to the shot noise of the probe. This proposal motivates a novel method for exploring thermal and quantum fluctuations in the solid state in a non-invasive manner. 
\end{abstract}

\maketitle

\section{Introduction}

\par Typically, noise is associated with unwanted part of the measurement, obscuring the signal under study. However, noise can also be a powerful metrological resource. For instance, measurements of spontaneous spin noise have been used as a probe of nanosecond spin dynamics and magnetic resonance in paramagnetic alkali atoms \cite{Crooker2004}, or proposed for finding the ground state in spin-glass systems \cite{Nishimura2016PRA}. In a pulsed regime, analysis of noise has helped to improve coherent Raman spectroscopy \cite{Xu2008NatPhys,Tollerud2019PNAS}. Even quantum noise of electromagnetic fields has been probed through matter-assisted transfer of polarization noise on the optical beam\cite{Riek2015Science}. Thus, noise spectroscopy is emerging as a crucial tool in condensed matter physics, which can provide deep insights into the intrinsic properties and dynamic behaviors of materials, devices and incoming fields. Unlike classical spectroscopy, which primarily focuses on the linear response of a system to external stimuli, noise spectroscopy investigates the fluctuations inherent to the system. Despite these successes, there has not been a general proposal for studying charge and spin noise in the solid-state, starting with the prototypical example of semiconductors. Here we show how a successful probe of fluctuations arising from stochastic displacement of charges can be used as a sensitive method to probe equilibrium and out-of-equilibrium thermal currents in a semiconductors. %That's why the interest of noise spectroscopy methods applied to semiconductors, fundamental materials to modern optoelectronics, rises, because that field of research gives us the ability to access optically in an non-invasive manner to the quantum behavior inside materials, such as the electron-hole recombination or tunneling effects.
\\
\par In particular, leveraging materials with dynamic optical properties under the influence of applied DC currents enables the extraction of quantum properties while controlling the bias. Recent advancements in electro-optic sampling schemes\cite{eos} and spectroscopies based on the Pockels effect demonstrate sensitivity to internal DC field, and offer a nuanced approach to probing quantum phenomena. Besides, we figure out that for metals or highly conductive semiconductors far more sensitive to current fluctuations than electric field fluctuations, a method where the optical signal at the ouput of the crystal relying more on charge or spin currents rather than electric field would be more sensitive. It motivated us to exploit the generation of second harmonic thanks to the presence of a direct electric current within a semiconductor, a phenomenon theoretically demonstrated in the 90's and known as current-induced second harmonic generation (CSHG)\cite{SHGTh}. An applied DC current breaks the symmetry of charge in space, which entails an assymetry of the electron and hole distribution in momentum space. The latter couples with the incident first harmonic to generate a second harmonic, with an effective second order non-linear susceptibility $\vb{\chi}_J^{(2)}$ proportionnal to the current $\vb{J_{\mathrm{DC}}}$
\begin{equation}\label{chi2eff}
    \vb{\chi}_J^{(2)}=\vb{\chi}_{\mathrm{c}}^{(3)} \vb{J_{\mathrm{DC}}},
\end{equation} with $\vb{\chi}_{\mathrm{c}}^{(3)}$ an effective susceptibility induced by the current $\vb{J}_\textrm{DC}$\cite{SHGTh}. the The proportionality relation \ref{chi2eff} has already been demonstrated experimentally with a DC current applied in generic centrosymmetric semiconductors, for instance composed of Silicon\cite{Si} or in 100-oriented Gallium Arsenide (GaAs)\cite{SHGExp} with a cautious experimental method to discriminate it with the SH generated. 
\\
\par Now, we shift from the DC picture to the quasi-stationnary picture. Indeed, the advent of femtosecond lasers generating ultra-short pulses pave the way to the time-resolved noise spectroscopy of short transient processes, like ultrafast charge transport. On ultrashort time scales, experimental physicists can freeze quantum fluctuations into essentially statistical realizations of systems under perturbatively small fluctuating fields. Noise spectroscopy makes the most of ultrafast probe seeing these fluctuating fields as quasi-stationary fields, and these techniques may shed light on ultrafast carrier dynamics in semiconductors, which is essentially our paper goal. 
\\
\par In thermal equilibrium, random motion of free carriers in a semiconductor largely stems from thermal agitation which dephase on ultrafast timescales of few- to sub-picoseconds, depending on temperature. In quantum matter, such fluctuations might also be adorned with quantum correlations, underpinning the formation of a macroscopic quantum state. As time relaxation of free carriers in a semiconductor is very fast, averaging the thermal current over the nanosecond ($10^{-9}\;\mathrm{s}$) with modern photodetectors makes it null. Our approach relies on the probe of ultrafast electronic transport with pulsed light possessing a characteristic duration far below the time characteristic of these currents, at such short time scales that the electron thermal motion appears as a frozen coherent current. So here we propose an optical probe of the microscopic electrical current fluctuations exploiting the third order nonlinear optical process mentioned above. We want to see if second harmonic signals driven by fluctuating stochastic currents are detectable. 
\\
\par Even though this paper is essentially theoretical, we have at disposal in our laboratory femtoQ a femtosecond laser PHAROS coupled to an Optical Pulse Amplifier (OPA) ORPHEUS (both designed by \textit{Light Conversion}, which make possible to probe optically the microscopic thermal current according to the setup proposed.

\section{\label{sec2}Theory}

\subsection{\label{2.A}Quadatric optical response for a centrosymmetric semiconductor}

Consider an incident electric field $\vb{E_\mathrm{in}}(\vb{r},t)$ and a centrosymmetric crystal, so that $\vb{\chi^{(2)}}_{\mathrm{bulk}}=0$. We make the following assumptions: 
\begin{itemize}
    \item We study light-matter interaction in the dipolar approximation. In addition, the layer of doped material is thin enough that we can abandon all spatial dependency.
    \item We neglect the field dispersion inside the material.%(Jacob)Why is that important?
    \item  We study the free carriers dynamics inside the material (bulk) and neglect surface contributions. The dependency of the latter to the beam input angle differ from that of the bulk, and there usually exist experimental conditions where they can be safely neglected.
\end{itemize}

Second-harmonic generation in a centro-symmetric system requires a break in the symmetry of space. In our analysis, this symmetry break is provided by an external DC electric field. Under these conditions, regular four-wave mixing may be provided with the DC field, leading to electric-field induced second harmonic generation (EFISH)\cite{EFISH}. In addition, the DC field can induce a current in sufficiently conductive samples, and that current in turn acts as an effective second-order susceptibility term, and separate from the regular nonlinear susceptibilities of the material. Same happens when the current is injected balistically. This leads to CSHG~\cite{SHGTh}. Yet, we don't apply any DC current nor DC field in our spectroscopy. 
\\
\par
\begin{figure}
    \centering
    \includegraphics[width=\linewidth]{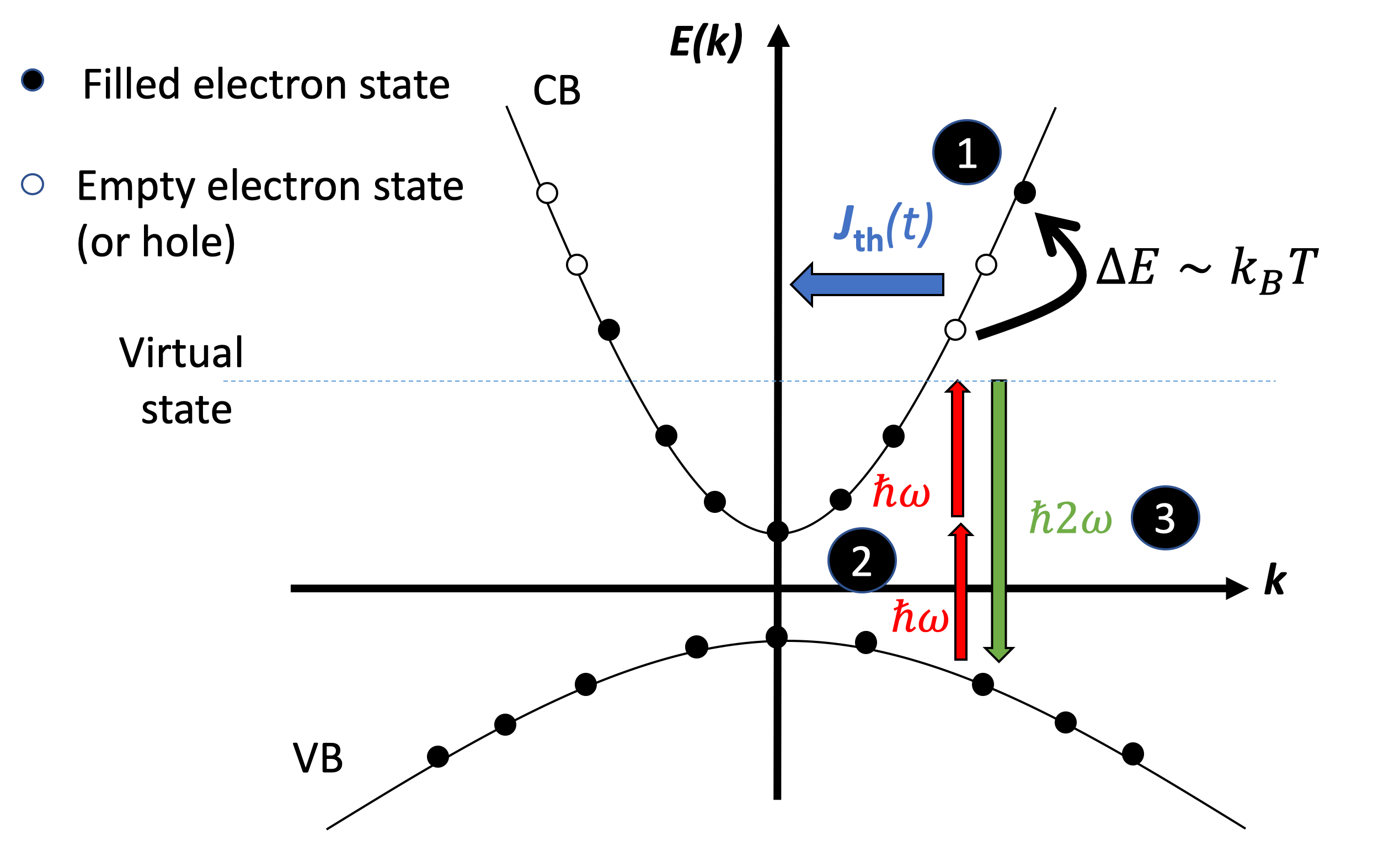}
    \caption{Simplified band diagram of the states involved in the TCSHG. The sequence of events leading to TCSHG is: 1. At any given time, thermal agitation breaks the symmetry of repartition in momentum of the free carriers, generating a thermal current flucutating at a sub-picosecond timescale 2.A two-photon transition to a virtual state caused by the absorption of two incident photons at $\omega$ 3.Spontaneous intraband transition induces a second-harmonic polarization and causes emission of a SH photon.}
    \label{banddiagram}
\end{figure}

 Thermal agitation causes spontaneous asymetry in the repartition momentum of electrons in the conduction band, as highlighted in figure \ref{banddiagram}, and generates a fluctuating thermal current $\vb{J_{\mathrm{th}}}$. In the presence of an incoming field at angular frequency $\omega$ with a femtosecond pulse, thermal current acts as a DC current and the nonlinear polarization of the material at angular frequency $2\omega$ is thus (see proof in \ref{polarizationproof})
\begin{equation}
\label{polarization2omega}
    \vb{P}(2\omega)= \epsilon_0\left[\chi^{(3)}_\textrm{m}\,\vb{E}_\textrm{th}\,+\chi_\mathrm{c}^{(3)}\,\vb{J}_\textrm{th}\right]\,\vb{E}^2_\textrm{in}(\omega),
\end{equation}
where $\vb{\chi}_\textrm{m}^{(3)}$ is the regular third-order susceptibility of the material and $\vb{E}_\textrm{th}$ the electric field due to electrons, considered as non-resonant terms. This "screening" field may contribute to the semiconductor total response for optical two-photon transitions below the band gap. So here, and in sufficiently conductive samples of GaAs, the current-induced term largely dominates, allowing us to neglect the EFISH contribution in our analysis.

\subsection{Thermal noise sensing}

At a given equilibrium temperature, we assume for a random thermal current $\vb{J}_{\mathrm{th}}$ a Maxwellian distribution  . Then, thermal agitation induces Johnson-Nyquist noise with the following statistics $\ev{\vb{J}_{\mathrm{th}}}=0$ and $\Delta\vb{J}_{\mathrm{th}}\equiv\sqrt{\ev{\vb{J}^2_{\mathrm{th}}}}=\sqrt{\frac{4k_B\,T\sigma\,B}{V}}$, where $k_B$ is Boltzmann's constant, $T$ the temperature, $\sigma$ the conductivity of the material, $B$ the detection bandwidth, and $V$ the sampled volume.
\\
\par Under the application of an incident electric field at angular frequency $\omega$, the thermal current will induce a random effective polarization of the material at angular frequency $2\omega$, with the following statistical characteristics: $\ev{\vb{P}_{\mathrm{th}}(2\omega)}=0$ and $\Delta\vb{P}_{\mathrm{th}}(2\omega)=\epsilon_0\chi_\textrm{c}\;\Delta \vb{J}_{\mathrm{th}}$. This polarization term will then induce an output field at angular frequency $2\omega$, which we term thermal-current-induced second-harmonic generation (TCSHG, see Figure \ref{banddiagram}).
\\
\par For an isotropic media in thermal equilibrium, the equipartition theorem implies that thermal current  components have the same distribution in all directions.
Here, we treat the co-polarized case where the pump propagates along $\vb{e_z}$ and the electric field polarized along $\vb{e_x}$.
We consider that only the parallel component of the thermal current couples to the incident field and generates a signal polarized along $\vb{e_x}$.
We take the value of $\chi_\textrm{c}$ to be the experimentally measured~\cite{SHGExp} $1\times10^{-22}\;\textrm{m}^3/\textrm{W}$.

\subsection{Time scale dynamics}

The free carriers momentum relaxation time $\tau_{J}$, or scattering time, characterizes the mean time between two events which will make the free carrier (electron or hole) change its direction and/or its velocity. For a given semiconductor, it depends only on the thermal energy $k_\mathrm{B}T$, so the frequency range $B_\mathrm{J}$at room temperature of the transient current lies in $\frac{k_\mathrm{B} T}{2\pi\hbar} \sim \num{1}\unit{\tera\hertz}$ \label{BJ}. Equivalent to say time scale of thermal current fluctuations lies in sub- or picoseconds. This order of magnitude is consistent with many pump probe measurements\cite{pumpprobe}.  Hence we can see the current as essentially frozen with a pulse width $\tau$ of the order of dozens of femtosecond.\\

\par \label{rectime} The recombination time $\tau_r$, which characterizes an exciton lifetime in the semiconductor after its interaction with light, is of the order of the nanosecond. In other words, promoted electrons by photon absorption to the conduction band go back to their ground states in the valence band in nanosecond timescale, much greater than the relaxation time but much smaller than the typical repetition time of lasers. Then, the information carried by each pulse is independent from one another.

\subsection{Measurement method}

\par The repetition frequency of our pulsed laser can be set at $\Delta f = 100 \;\mathrm{kHz}$, i.e a period of $10^{-5}\;\mathrm{s}$, much smaller than the recombination time scale. We can assume then that the information contained in the output of each pulse is independent of that of other pulses. We define the statistic average of a physical quantity $X$ 
\begin{equation}
    \overline{X}:= \frac{1}{N_{\mathrm{pulses}}} \sum_{\mathrm{pulses}} \expval{X}(p\Delta t),
\end{equation}
where $N_{pulses}$ is the number of pulses under a given time interval T. The ergodicity hypothesis entails $\Delta^2 J_{\mathrm{th}}=\overline{J^2_{\mathrm{th}}}$. This identification is the keystone of our time-resolved noise spectroscopy.

\section{Thermal current induced second harmonic generation (TCSHG)}

\subsection{Estimation of thermal current fluctuations}
\par Usually, energy states of opposition momentum in $\vb{k}$-space generates SH polarization with opposite signs. But in the case where spontaneous intraband transitions happen due to thermal agitation as illustrated in Figure \ref{banddiagram}, there is a chance that state +$\vb{k}$ is blocked for SH generation (either electron in CB or hole in VB) while -$\vb{k}$ is available (either hole in CB or electron in VB). Without loss of generality, we can estimate $\Delta{J_{\mathrm{th}}}$ when current fluctuations arise due to thermal equilibrium of GaAs semiconductor at 300 K, once we assumed Maxwellian distribution at a given equilibrium temperature. The different experimental values concerning material properties or laser parameters have to be set to enhance thermal noise. As described in appendix \ref{estimdeltaJ}, we minimize the volume where light and matter interacts and enhance considerably the carriers number of that volume. Numerically, thermal current fluctuations amount to
 \begin{equation}
     \Delta\;J_{\mathrm{th}} \sim 3 \times 10^6 \;\unit{A.m^{-2}}.
 \end{equation}

\subsection{TCSHG power}

For an input laser power $P_{\mathrm{Laser}}(t)$, the optical power of the TCSHG beam at the output of the crystal writes
\begin{equation}
    P_{\mathrm{Signal}}(t)=\Lambda_{\mathrm{Signal}} P_{\mathrm{Laser}}^2(t)J_{\mathrm{th}}^2(t),
\end{equation}
with $\Lambda_{\mathrm{Signal}}:=\frac{2|\eta_{\mathrm{Signal}}|^2}{Sc\epsilon_0}$ and 
$\eta_{\mathrm{Signal}}=\frac{i\omega_c\chi_{\mathrm{c}} l}{4nc}$,
where n is the refractive index of the centrosymmetric conductor. Yet, the contribution of thermal fluctations to the signal is extremely weak: $\Delta^{\mathrm{th}} p_{\mathrm{signal}}\approx \num{13}\unit{\micro\ampere}$(see appendix \ref{undetect}). The information on thermal current statistics will be drowned out by the thermal noise of photodetectors. That's why we propose to amplify thermal current contribution to photocurrent fluctuations by means of a local oscillator in balanced homodyne scheme.

\section{Balanced Homodyne detection}

\subsection{Proposed experimental setup}
\begin{figure}
    \centering
    \includegraphics[width=\linewidth]{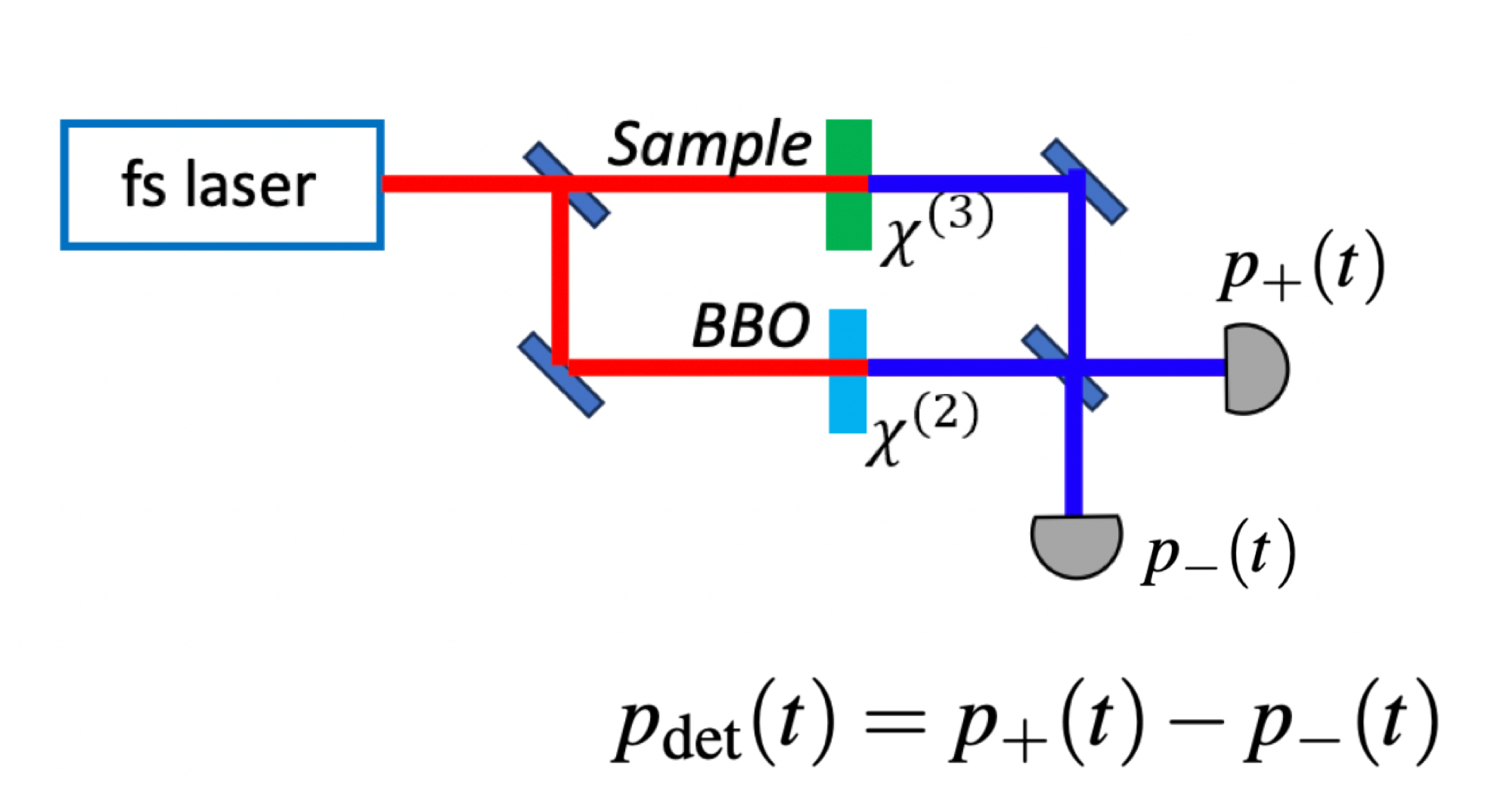}
    \caption{Balanced homodyne detection for measuring thermal current statistics. $p_{\mathrm{det}}(t)$ is the analysed temporal photocurrent, corresponding to the difference of the two photocurrents $p_+(t)$ and $p_-(t)$ detected at each arm of the interferometer.}
    \label{setup}
\end{figure}

The setup entails a sub-picosecond excitation of matter at frequencies below the band edge, resulting in production of second harmonic light via a four-wave mixing with a spontaneous intraband polarization. The local oscillator (LO) signal for homodyne amplification is then derived from frequency doubling (SHG) of portion of the excitation light, resulting in a linear interference of the thermal current-induced second harmonic generation (TCSHG) and the x-polarization of the LO signals. Each beam divided by the beamsplitter is collected with a photodetector, and the difference of the two is temporally analysed with an oscilloscope. The whole setup is presented in \ref{setup}.

\par Balanced homodyne allows us to get rid of inherent fluctuations of the laser intensity and of the dark current (noise detector) to enhance significantly the sensitivity on photocurrent detected fluctuations linked to thermal current fluctuations. Yet, a shot noise due to the local oscillator adds to the thermal noise steming from TCSHG. We want to increase the latter with regard to the shot noise.
\subsection{Probe and LO combining}
With a proper propagation through optics, each beam before recombination can be associated to cylindric beam of radius $\frac{w}{2}$.
LO and signal mixing leads to a cross optical interference term, of optical power half of $P_{\mathrm{cross}}(t)\equiv\Lambda_{\mathrm{cross}}P^2_{\mathrm{Laser}}(t)J_{\mathrm{th}}(t)$, with $\Lambda_{\mathrm{cross}}=\frac{2\sqrt{S_0}}{c\epsilon_0\sqrt{S^3}}|\eta_{\mathrm{LO}}\eta_{\mathrm{Signal}}|$.

\subsection{Detected photocurrent }
The detected photocurrent  $p_{\mathrm{det}}(t)$ equals to the cross photocurrent (see appendix \ref{photocurrentdetected}). At times where the photocurrent reaches its peak, it expresses $p_{\mathrm{det}}(t)=\Lambda^{\mathrm{det}}_{\mathrm{cross}}P^2_{\mathrm{Peak}}J_{\mathrm{th}}(t)$, with $\Lambda^{\mathrm{det}}_{\mathrm{cross}}=\frac{\sqrt{\pi} R\tau}{2\sqrt{2\ln2}T_m}\Lambda_{\mathrm{cross}}$. 
\section{Predictive experimental results}

All the issues for this noise spectroscopy is to determine whether the thermal current fluctuations contribute enough to the total detected intensity fluctuations $\mathrm{\Delta^\mathrm{tot} \;p_{\mathrm{det}}}$, to estimate the feasibility of detecting fluctuations. So using this proposed scheme, we theoretically investigated the contributions of current fluctuations $\mathrm{\Delta^\mathrm{th} \;p_{\mathrm{det}}}$ to the total detected intensity fluctuations $\mathrm{\Delta^\mathrm{tot} \;p_{\mathrm{det}}}$, assuming the latter are at the standard quantum limit. This allows us to write down a general expression for the ratio of thermal fluctuations over the shot noise $\mathrm{\Delta^\mathrm{SN} \;p_{\mathrm{det}}}$ (dominated by the LO), noted ToS.

\subsection{Detected photocurrent fluctuations}
The thermal contribution to the total fluctuations is $\Delta^{\mathrm{th}}p_{\mathrm{det}}=\Lambda^{\mathrm{det}}_{\mathrm{cross}}P^2_{\mathrm{Peak}}\Delta J_{\mathrm{th}}$, while the shot noise amounts to $\Delta^\mathrm{SN} p_{\mathrm{det}}=\frac{\sqrt{2}}{2}\Delta^\mathrm{SN} p_{\mathrm{LO}}$. The numerical values use not already mentioned are $\Lambda^{\mathrm{det}}_{\mathrm{cross}} = \num{1.2d-25}\unit{\square\meter\per\square\watt}$ and $P_{\mathrm{Peak}}= \num{1.4e7}\unit{\watt}$.
\\
\par Finally, we obtain numerically 
\begin{equation}
    \left\{\begin{aligned}
        &\Delta^\mathrm{SN} p_{\mathrm{det}}=\num{0.92}\unit{\milli \ampere}\\
        &\Delta^\mathrm{th} p_{\mathrm{det}}=68\unit{\micro\ampere}\\
    \end{aligned}
    \right.
\end{equation}
\\
\subsection{Thermal over Shot Noise ratio as a function of $P_{\mathrm{peak}}$ and $\Delta J_{\mathrm{th}}$}

The main result of our paper is the linear relationship between the ratio of thermal noise over shot noise for the detected intensity, noted ToS, with the product of the incident optical intensity $P_{\mathrm{Peak}}$  by the standard deviation of thermal current gaussian distribution $\mathrm{\Delta J_{\mathrm{th}}}$

\begin{equation}
    \mathrm{ToS}\equiv \frac{\Delta^\mathrm{th} \;i_{\mathrm{det}}}{\Delta^\mathrm{SN} \;i_{\mathrm{det}}}\equiv \kappa_{\mathrm{ToS}} P_{\mathrm{Peak}} \Delta J_{\mathrm{th}},
\end{equation}
with $\kappa_{\mathrm{ToS}}\equiv \frac{\Lambda^{\mathrm{det}}_{\mathrm{cross}}}{\kappa_{\mathrm{LO}}} = \num{2.1d-16}\unit[per-mode = fraction]{\raiseto{4}\meter\per\ampere \per \watt}$. In particular, the noise contribution evolves linearly with the number of photons, and is inversely proportional to the spot size of the beam, incident on the material. 
\par As $\Delta^\mathrm{tot} \;p_{\mathrm{det}}=\sqrt{1+\mathrm{ToS}^2}\;\Delta^\mathrm{SN} \;p_{\mathrm{det}}$, and using optical parameters of the laser and material properties, we estimate that in-quadrature addition of both noise terms amounts to 8\% additional noise contribution, detectable in suitably prepared laboratory environments.
\section{Conclusion}

 We propose a novel scheme for providing all optical access to intraband current fluctuations in materials, and estimate a few-percent coupling for thermal fluctuations in exemplary semiconductors, such as GaAs, which require precise experimental conditions for a sensitive sensing \cite{Riek2015Science}. This technique requires no contact nor bias fields, and hence is poised to provide novel experimental method for studying thermal as well as quantum fluctuations in solid state in a non-invasive manner. We work out here a classical case to see if our technique enables to see small fluctuations, so we envision that our approach can lead to a new modality of noise spectroscopy adaptible for studying quantum fluctuations in condensed matter in and out-of-equilibrium. Hence, these calculations guide experimental design for implementation of the scheme, underway in our laboratory.

\appendix

\section{Polarization at $2\omega$ induced by the thermal current}
\label{polarizationproof}
We consider a frequency $\omega$ in the incident pulse bandwidth. The third $\vb{\chi_{\mathrm{c}}^{(3)}}$ and second $\vb{\chi_J^{(2)}}$ order effective susceptibilities are linked through
\begin{equation}
    \vb{\chi_J^{(2)}}(2\omega;\omega_1,\omega_2)  \equiv \int_{\mathbb{R}} \vb{\chi_{\mathrm{c}}^{(3)}}(2\omega;\omega_1,\omega_2,\Omega') \tilde{J_{\mathrm{th}}}(\Omega') \;\frac{\dd\Omega'}{2\pi}
\end{equation}
As $J_{\mathrm{th}}$ can be treated as a DC current, with a spectrum at the bandedge of the incident light spectrum \begin{equation}
    \chi_J^{(2)}(2\omega;\omega_1,\omega_2)\approx  \chi_{\mathrm{c}}^{(3)}(2\omega;\omega_1,\omega_2,\Omega) \tilde{J_{\mathrm{th}}}(\Omega),
\end{equation}
with $\Omega$ in the THz range, as the transient current had an unique frequency component of value quasi-null.\\
\par We assume next a dispersionless media: the $\vb{\chi_{\mathrm{c}}^{(3)}}$ tensor is then independent from frequency. It comes in the co-polarized case when considering only a single component of $J_{\mathrm{th}}$
\begin{equation}
    \chi_J^{(2)}  \equiv \chi_{\mathrm{c}}^{(3)} J_{\mathrm{th}}(\Omega).
\end{equation}
\\
\par Therefore, in the co-polarized case we move to scalar notations to write the response of the centrosymmetric crystal
\begin{equation}
    \begin{aligned}
        P^{(3)}(t)\equiv&\epsilon_0 \int_{\mathbb{R}^2} \chi_J^{(2)}(\omega_1,\omega_2) \tilde{E_{\mathrm{in}}}(\omega_1)\tilde{E_{\mathrm{in}}}(\omega_2)\\&e^{-i(\omega_1+\omega_2)t} \frac{\dd\omega_1}{2\pi}\frac{\dd\omega_2}{2\pi}\\
        =&\epsilon_0 \int_{\mathbb{R}^3} \chi_{\mathrm{c}}^{(3)}(\omega_1,\omega_2,\Omega) \tilde{E_{\mathrm{in}}}(\omega_1)\tilde{E_{\mathrm{in}}}(\omega_2)e^{-i(\omega_1+\omega_2)t}\\&\tilde{J_{\mathrm{th}}}(\Omega)e^{-i\Omega t} \frac{\dd\omega_1}{2\pi}\frac{\dd\omega_2}{2\pi}\frac{\dd\Omega}{2\pi}\\
        =&\epsilon_0 \chi_{\mathrm{c}}^{(3)} J_{\mathrm{th}}(t)\int_{\mathbb{R}^2}  \tilde{E_{\mathrm{in}}}(\omega_1)\tilde{E_{\mathrm{in}}}(\omega_2)e^{-i(\omega_1+\omega_2)t}\frac{\dd\omega_1}{2\pi}\frac{\dd\omega_2}{2\pi}\\
        =&\epsilon_0  \chi_{\mathrm{c}}^{(3)} \,E^2_{\mathrm{in}}(t)\,J_{\mathrm{th}}(t).
    \end{aligned}
\end{equation}
%(Jacob)provide the fromula for xhi^3...or at elast relate it to the index nonlienarity 
\section{Choice for the source of light and the photodetector}

In our noise spectroscopy, the probe is a femtosecond pulse with a central frequency $\omega_c$ such that the two-photon excitation is nearly resonant with the semiconductor gap. The beam is supposed gaussian. The electric field at the ouput of the laser for a pulse in time domain, when the mean amplitude over the cross section of the pulse equals to $E_{\omega_c}$, expresses
\begin{equation}
    E_{\mathrm{Laser}}(t)=E_{Laser}e^{i\omega_c t} e^{-\Gamma t^2}.
\end{equation}
The characteristic time between the beginning and the end of the pulse is called pulse duration, and is defined by the FWHM (full width at half maximum) of the intensity shape
\begin{equation}
    \tau=\sqrt{\frac{2\ln2}{\Gamma}}.
\end{equation}
We want a quasi-resonant interaction, roughly 30\% above the band gap. At 300K, for GaAs, the band gap is 1.422 eV, so we need to reach 1.85eV, which corresponds to a wavelength of 672nm. The two incident photons would have then 1.3 micrometers as wavelengths each. This thought process leads us to choose as incident wavelength (or equivalently pulsation):
\begin{equation}
    \left\{\begin{aligned}
    \lambda_c =& \num{1.3}\unit{\micro\meter}\\
    \omega_c= &\num{1.5e15}\unit{\per\second}
    \end{aligned}\right.
\end{equation}

\par The predictive results of our proposal are based on ORPHEUS-F (from \textit{Light Conversion}) specs for such a $\lambda_c$, whose parameters are the average output power $P_{\mathrm{avg}}= \num{0.7}\unit{W}$, the pulse width $\tau = \num{50}\unit{\femto\second}$, the repetition frequency $f_{\mathrm{rep}}=1\unit{MHz}$ and the beam diameter at the output of the laser: $w\sim \num{1}\unit{\milli\meter}$. The energy per pulse is then $E_p = \num{0.7}\unit{\micro\joule}$ and the peak optical power, i.e the maximum instantaneous optical power emitted by the laser, is $P_{\mathrm{Peak}}=\num{1.4e7}\unit{\watt}$.

\par After the laser beam propagation through a suitable collimation followed by a suitable focus, we coïncide the beam waist with the sampled volume where the $2\omega$ polarization occurs. By energy conservation, $E_{\mathrm{in}}= \sqrt{\frac{S}{S_0}}E_{\mathrm{Laser}}$, and the incident field writes then
\begin{equation}
     E_{\mathrm{in}}(t)=E_{\mathrm{in}}e^{i\omega_c t} e^{-\Gamma t^2},
\end{equation}
where $S$ and $S_0$ the beam cross section at respectively the laser output and at the beam waist. The TCSHG beam, or signal, diameter remains $w_0$ at the crystal output, with an instantaneous optical power $P_{\mathrm{Signal}}$, which writes
\begin{equation}
    P_{\mathrm{Signal}}(t)= S_0 I_{\mathrm{Signal}}(t)
\end{equation}

\par We examine then how the signal beam light gets turned into a photocurrent. To do so, we need to introduce generic photodiode characteristics suitable for our experimental proposal: a response time of $T_m \sim \num{10}\unit{\nano\second}$ and a photodiode responsivity of $R\sim \num{1} \unit{\ampere\per\watt}$ (linked to the quantum efficiency $\eta$). The photocurrent $p(t)$ is derived from the integration of the instantaneous optical power $P_{\mathrm{Signal}}(t)$ during the interval of time $T_m$ via
\begin{equation}
    p(t)=R \expval{P_{\mathrm{Signal}}} (t).
\end{equation}

\section{Estimation of thermal current fluctuations}
\label{estimdeltaJ}
\par First, we have to define the volume V under study, which is defined by the product of the characteristic length of the light/matter interaction in the semiconductor $\ell$ with the beam waist $w_0$. Smaller the volume is, bigger the thermal fluctuations are. So we aim to minimize both $\ell$ and $w_0$. At $\num{1300}\unit{\nano\meter}$, the absorption depth is on the order of cm. So we can heavily dop a thin layer of $\num[]{1}\unit{\micro \meter}$ of length. As for the beam waist, it is only limited by the diffraction, so we fix $w_0=\num{50} \unit{\micro\meter}$. Then, a volume of $V \sim \num{2e3} \unit{\micro\cubic\meter}$.
\\
\par We've seen in \ref{BJ} that $B_J \sim 1 \; \mathrm{THz}$. That's consistent with a DC current: the thermal current distribution in frequency is very narrow and around the input pulsed beam band edge on the order of hundreds of THz.
\\
\par Thirdly, in a n-doped semiconductor the conductivity is function of free electrons density and mobility as $\sigma_{sc}=e n_e\mu_e $ with $n_e$ the electron density and$\mu_e$  the electron mobility. The litterature \cite{SHGTh,SHGExp} teaches us we need a heavily dop crystal to see current effects, with carriers density up to $\num{1e19}\unit{\cubic\centi\meter}$. Thus the conductivity of the centrosymmetric crystal can be raised to
$\sigma_{sc} \sim 10^{6} \unit{S.m^{-1}}.$

\section{Contribution of thermal current fluctuations to the photocurrent without a balanced homodyne detection scheme } \label{undetect}

With regard to the polarization at $2\omega$ induced within the material \ref{polarization2omega}, the expression for the second harmonic field generated at the output of the semiconductor with refractive index n and characteristic length $\ell$ is 
\begin{equation}
     E_{\mathrm{Signal}}(t)=\eta_{\mathrm{Signal}}\,E_{\mathrm{in}}^2(t)\,J_{\mathrm{th}}(t),
\end{equation}
with $\eta_{\mathrm{Signal}}:= i\frac{\chi_{\mathrm{c}}\omega_c \ell}{4nc}$. Noting $\Lambda_{\mathrm{Signal}}:=\frac{2n|\eta_{\mathrm{Signal}}|^2}{Sc\epsilon_0}$, the corresponding optical power writes
\begin{equation}
    P_{\mathrm{Signal}}(t)=\Lambda_{\mathrm{Signal}} P_{\mathrm{Laser}}^2(t)J_{\mathrm{th}}^2(t),
\end{equation}

With a photodetector at the output of the crystal, we detect
\begin{align*}
    p_{\mathrm{Signal}}(t)=&R\expval{ P_{\mathrm{Signal}}(t)}\\
 =&R\Lambda_{\mathrm{Signal}} \expval{P_{\mathrm{Laser}}^2(t)\;J_{\mathrm{th}}^2(t)}.
\end{align*}
with \mbox{$\expval{.}$} defined as a function of time by

\begin{equation}
    \expval{P_{\mathrm{Signal}}} (t)= \frac{1}{T_m} \int_{t-\frac{T_m}{2}}^{t+\frac{T_m}{2}} \mathrm{d}t' \;P_{\mathrm{Signal}}(t').
\end{equation}

As $\tau_J \gg \tau $, we can consider the thermal current as "frozen" during the interaction between the light pulse and the crystal. Everything happens as if the incident electric field saw a DC thermal current during the time of coupling, which allows to sample the transient thermal current over its values at the peak pulse. For $p\in\mathbb{N}$:
\begin{equation*}
\expval{P_{\mathrm{Laser}}^2(t)\;J_{\mathrm{th}}^2(t)} = \expval{P_{\mathrm{Laser}}^2(t)}J_{\mathrm{th}}^2(p\Delta t)
\end{equation*}

Incidentally,
\begin{equation}
\begin{aligned}
    \expval{P_{\mathrm{Laser}}^2(t)}=&\frac{1}{T_m}P^2_{\mathrm{Peak}}\int_{\mathbb{R}}\;\mathrm{d}t\;e^{-2\Gamma t^2}\\
    =&\frac{\sqrt{\pi}\tau}{2\sqrt{\ln2}T_m}P^2_{\mathrm{Peak}}
\end{aligned}
\end{equation}

Eventually, noting $\Lambda^{\mathrm{det}}_{\mathrm{Signal}}=\frac{R\sqrt{\pi}\tau}{2\sqrt{\ln2}T_m} \Lambda_{\mathrm{Signal}}$, the TCSHG photocurrent expresses
\begin{equation}
    p_{\mathrm{Signal}}(p\Delta t)= \Lambda^{\mathrm{det}}_{\mathrm{Signal}} \;P_{\mathrm{Peak}}^2  \;J_{\mathrm{th}}(p\Delta t)^2,
\end{equation}

TCSHG is already thermal noise, because induced by thermal current fluctuations. So the thermal contribution to fluctuations is just the mean square root of TCSHG photocurrent. Noting $\kappa^{\mathrm{th}}_{\mathrm{Signal}}= \sqrt{2}\Lambda^{\mathrm{det}}_{\mathrm{Signal}}$, we obtain

\begin{equation}
\begin{aligned}
    \Delta^\mathrm{th} p_{\mathrm{signal}}\equiv &\sqrt{\overline{p_{\mathrm{signal}}^2}-\overline{ p_{\mathrm{signal}}}^2}\\
=&\kappa^{\mathrm{th}}_{\mathrm{TCSHG}}\;P_{\mathrm{Peak}}^2\;\Delta J_{\mathrm{th}}^2.
    \end{aligned}
\end{equation}

\par We estimate the magnitude of the thermal noise with experimental values for GaAs, where our reference \cite{SHGExp} gives $\chi_{\mathrm{c}}^{(3)} \sim 10^{-22} \;\unit{m^3 W^{-1}}$.
A numerical computation leads to $\Delta^\mathrm{th} \;p_{\mathrm{Signal}}=\num{1.3d-5}\unit{\ampere}$.
\par As for shot noise, $\Delta^{\mathrm{SN}} p_{\mathrm{Signal}}\equiv\sqrt{2eB_{\mathrm{Signal}}\overline{p_{\mathrm{Signal}}}}$. We note $\kappa^{\mathrm{SN}}_{\mathrm{Signal}} = 4\sqrt{\frac{e\ln2}{\tau}\Lambda^{\mathrm{det}}_{\mathrm{Signal}}}$. Eventually,
\begin{equation}
     \Delta^{\mathrm{SN}} p_{\mathrm{Signal}}=\kappa^{\mathrm{SN}}_{\mathrm{Signal}}P_{\mathrm{Peak}}\Delta J_{\mathrm{th}}.
\end{equation}
\par Numerical computations give $\Delta^{\mathrm{SN}} p_{\mathrm{Signal}}=\num{2.1d-5}\unit{\ampere} \sim 2  \Delta^{\mathrm{th}}p_{\mathrm{Signal}}$.

\par At this stage, the total fluctuations equal to
 \begin{align*}
            \Delta^\mathrm{tot} p_{\mathrm{Signal}}=&\sqrt{(\Delta^\mathrm{th} p_{\mathrm{Signal}})^2+(\Delta^\mathrm{SN} i_{\mathrm{Signal}})^2}\\
            \approx & \num{25}\unit{\micro\ampere},
\end{align*}
and so are hardly detectable in a mere optical configuration due to thermal noise of the photodetector.

\section{Focus on the LO}

The only source of noise for the LO stems from the laser. Our framework is based on the following approximations:
\begin{itemize}
    \item non-depleted pump approximation: the SG generated amplitude is much lower than $E_{\mathrm{in}}(t)$, such that we consider the latter remaining constant over the interaction length L.
        \item perfect phase matching; i.e $\Delta k= k(2\omega)-2k(\omega)=0$.
        \item $\chi^{(2)}(BBO)$ dispersionless over the input beam bandwidth.
\end{itemize}
\par Then, using a symmetric beamsplitter and noting $\eta_{\mathrm{LO}}=\frac{i\omega_c \chi_2(BBO) L }{4 n' c}$, the LO field expresses
\begin{equation}
    E_{\mathrm{LO}}(t) = \frac{1}{2}\eta_{\mathrm{LO}}\,E_{\mathrm{in}}^2(t),
\end{equation}
where n' corresponds to the frequency-dependent BBO refractive indix and satisfies to
$n' \equiv n'_{BBO}(2\omega_c)=n'_{BBO}(\omega_c)$. We note then $\Lambda_{\mathrm{LO}}=\frac{n'|\eta_{\mathrm{LO}}|^2}{2Sc\epsilon_0}$. The LO power writes
\begin{equation}
    P_{\mathrm{LO}}(t)=\Lambda_{\mathrm{LO}} \;P^2_{\mathrm{Laser}}(t).
\end{equation}

\par We note $\Lambda^{\mathrm{det}}_{\mathrm{LO}}= \frac{R\sqrt{\pi}\tau}{2\sqrt{2\ln2}T_m}\Lambda_{\mathrm{LO}}$. When the pulse reaches its peak, it converts into a photocurrent 
\begin{equation}
    p_{\mathrm{LO}}(p\Delta t)=\Lambda^{\mathrm{det}}_{\mathrm{LO}} P^2_{\mathrm{peak}}
\end{equation}

\par The shot noise amounts to 
\begin{align}
    \Delta^{\mathrm{SN}} p_{\mathrm{LO}}=&\sqrt{2eB_{\mathrm{LO}}\overline{p_{\mathrm{LO}}}}\\
    =&\kappa_{\mathrm{LO}} P_{\mathrm{Peak}},
\end{align}
with $B_{\mathrm{LO}}$ the LO signal bandwidth and $\kappa_{\mathrm{LO}}=4\sqrt{\frac{e\ln2}{\tau}\Lambda^\mathrm{det}_{\mathrm{LO}}}$.

\par With $L \sim 100 \;\unit{\mu m}$ and $\chi_2(BBO) \approx 4.4 \times 10^{-12} \;\unit{m V^{-1}},$ the shot noise amounts to $ \Delta^{\mathrm{SN}} p_{\mathrm{LO}}= \num{1.3}\unit{\milli\ampere}$.

\section{Recombining signal optical powers}
\label{photocurrentdetected}
The optical power at each arm after the second beamsplitter writes
\begin{equation}
\begin{aligned}
         P_+(t)=&\frac{1}{2}Sc\epsilon_0\Biggl|\frac{1}{\sqrt{2}}\left(\frac{1}{2}\sqrt{\frac{S_0}{S}}E_{\mathrm{Signal}}(t)+E_{\mathrm{LO}}(t)\right)\Biggr|^2\\
        =&\frac{1}{2}P_{\mathrm{LO}}(t) +\frac{1}{8}\frac{S_0}{S}P_{\mathrm{Signal}}(t)
         +\frac{1}{2}P_{\mathrm{cross}}(t).
    \end{aligned}
\end{equation}

\par We note  $\Lambda_{\mathrm{cross}}=\frac{2\sqrt{S_0}}{c\epsilon_0\sqrt{S^3}}|\eta_{\mathrm{LO}}\eta_{\mathrm{Signal}}|$ and the cross interference term of the optical power 
\begin{equation}
\begin{aligned}
    P_{\mathrm{cross}}(t)\equiv &\frac{1}{2}\sqrt{S_0S}c\epsilon_0 \Re{E_{\mathrm{LO}}E^\star_{\mathrm{Signal}}}\\
    =&\Lambda_{\mathrm{cross}}P^2_{\mathrm{Laser}}(t)J_{\mathrm{th}}(t),
    \end{aligned}
\end{equation}

\par Thus, the photocurrent detected at each arm is
\begin{equation}
    p_\pm(t)=\frac{1}{2}p_{\mathrm{LO}}(t)+\frac{1}{8}\frac{S_0}{S}p_{\mathrm{Signal}} \pm \frac{1}{2}p_{\mathrm{cross}}(t),
\end{equation}
and the photocurrent detected via our homodyne detection scheme is
\begin{equation}
    p_{\mathrm{det}}(t)=p_{\mathrm{cross}}(t),
\end{equation}
where
\begin{equation}
    p_{\mathrm{cross}}(t)=R\expval{P_{\mathrm{cross}}(t)}
\end{equation}
\par As already said, we can sample $J_{\mathrm{th}}$ at $p\Delta t$ for $p\in\mathbb{N}$. We reexpress 
\begin{equation}
    \expval{P_{\mathrm{cross}}}(p\Delta t) =\frac{1}{T_m} \Lambda_{\mathrm{cross}}J_{\mathrm{th}}(p\Delta t)\expval{P^2_{\mathrm{Laser}}}(p\Delta t)
    \end{equation}
Eventually, noting $ \Lambda^{\mathrm{det}}_{\mathrm{cross}}=\frac{\sqrt{\pi} R\tau}{2\sqrt{2\ln2}T_m}\Lambda_{\mathrm{cross}}$,
\begin{equation}
    p_{\mathrm{det}}(p\Delta t)=\Lambda^{\mathrm{det}}_{\mathrm{cross}}P^2_{\mathrm{Peak}}J_{\mathrm{th}}(p\Delta t)
\end{equation}

\par We compute now the shot noise, stemming from the LO. By additivity of independent cumulants, shot noise at each arm equals to
\begin{equation}
\begin{aligned}
    \Delta^\mathrm{SN} p_{+}=&\sqrt{\frac{1}{4}\left(\Delta^{\mathrm{SN}}p_{\mathrm{LO}})\right)^2+\frac{1}{64}\left(\Delta^{\mathrm{SN}}p_{\mathrm{th}})\right)^2+\left(\Delta^{\mathrm{SN}}p_{\mathrm{cross}})\right)^2}\\
        \approx &\frac{1}{2}\Delta^{\mathrm{SN}}p_{\mathrm{LO}}
    \end{aligned}
\end{equation}
So the shot noise at the readout of the balanced homodyne detection is
\begin{equation}
    \begin{aligned}
         \Delta^\mathrm{SN} p_{\mathrm{det}}=&\sqrt{\left(\Delta^\mathrm{SN} p_{+}\right)^2+\left(\Delta^\mathrm{SN} p_{-}\right)^2}\\
            =&\frac{\sqrt{2}}{2}\Delta^\mathrm{SN} p_{\mathrm{LO}}.\\
    \end{aligned}
\end{equation}

\section*{Acknowledgement}
This work has been co-funded by the European Union's Horizon Europe Research and Innovation Programme under agreement 101070700 (project MIRAQLS).

\bibliography{mybib}% Produces the bibliography via BibTeX.

\end{document}